\documentclass[10pt,conference]{IEEEtran}
\usepackage{graphicx} %
\usepackage{amsthm}

\makeatletter
\@namedef{ver@lineno.sty}{9999/12/31}
\@namedef{opt@lineno.sty}{}
\makeatother
\usepackage[frozencache=true]{minted}

\usepackage{caption}
\usepackage{subcaption}
\usepackage{algpseudocode}
\usepackage{csvsimple}
\usepackage{varwidth}
\usepackage{pgfplots}
\usepackage[utf8]{inputenc}
\usepackage{newunicodechar}
\usepackage{upquote}
\usepackage{environ}
\usepackage[numbers]{natbib}
\usepackage{amsmath}
\usepackage{amsfonts}
\usepackage{amssymb}
\usepackage{booktabs}
\usepackage{cprotect}
\usepackage{pifont}%
\newcommand{\cmark}{\ding{51}}%
\newcommand{\xmark}{\ding{55}}%
\usepackage{xurl}
\newunicodechar{↩}{\ensuremath{\hookleftarrow}}
\usepackage{balance}
\usepackage{todonotes}

\setminted{bgcolor=yellow!10}

\pgfplotsset{compat=1.18}

\newtheorem{definition}{Definition}

\newtheorem{problem}{Problem}

\algblockdefx[Branch]{Branch}{EndBranch}[1]{\textbf{create branch #1; if in new branch then}}[0]{\textbf{end if}}
\algdef{SE}[SUBALG]{Indent}{EndIndent}{}{\algorithmicend\ }%
\algtext*{Indent}
\algtext*{EndIndent}

\date{}

\begin{document}

\newcommand{\lexemeAlphabet}{\Sigma_{\mathcal{G}}}
\newcommand{\charAlphabet}{\Sigma_{\mathcal{L}}}
\newcommand{\comments}{\Sigma_{\mathcal{C}}}
\newcommand{\lexemeLang}[1]{\mathcal{T}_{#1}}
\newcommand{\lexemeNFA}[1]{\mathcal{N}_{#1}}

\newcommand{\lex}{\textit{Lex}}

\newcommand{\nfaStateSpace}{Q}
\newcommand{\nfaAlphabet}{\Sigma}
\newcommand{\nfaInitialState}{s_0}
\newcommand{\nfaTransition}{\delta}
\newcommand{\nfaFinalStates}{F}

\newcommand{\nfaTuple}{(\nfaStateSpace, \nfaAlphabet, \nfaInitialState, \nfaTransition, \nfaFinalStates)}

\newcommand{\cfgNonterminals}{V}
\newcommand{\cfgRules}{R}
\newcommand{\cfgStart}{S}
\newcommand{\cfgAlphabet}{\Sigma}
\newcommand{\cfgTuple}{(\cfgNonterminals, \cfgAlphabet, \cfgRules, \cfgStart)}
\newcommand{\langOfCFG}[1]{L(#1)}

\newcommand{\reg}{\mathcal{R}}
\newcommand{\regsub}[2]{\reg[#1, #2]}

\newcommand{\cfl}{\mathcal{G}}
\newcommand{\cflsub}[1]{\cfl[#1]}
\newcommand{\textquote}{"}
\newcommand{\powerset}[1]{2^{#1}}
\newcommand{\pipe}{|}

\newcommand{\highlightCode}[1]{\begingroup\fboxsep=0pt\colorbox{green!25}{#1\strut}\endgroup}
\newcommand{\highlightCodeBlue}[1]{\begingroup\fboxsep=0pt\colorbox{blue!20}{#1\strut}\endgroup}
\newcommand{\highlightCodeRed}[1]{\begingroup\fboxsep=0pt\colorbox{red!20}{#1\strut}\endgroup}
\newcommand{\mintedComment}[1]{\begingroup\textit{\textcolor{gray}{#1}}\endgroup}

\newcommand{\sublang}[1]{\mathcal{L}'_{#1}}

\title
{Constrained Decoding for Fill-in-the-Middle Code Language Models via Efficient Left and Right Quotienting of Context-Sensitive Grammars}

\author{\IEEEauthorblockN{Daniel Melcer}
\IEEEauthorblockA{\textit{Northeastern University} \\
Boston, MA \\
melcer.d@northeastern.edu}
\and
\IEEEauthorblockN{Nathan Fulton}
\IEEEauthorblockA{\textit{MIT-IBM AI Lab} \\
Boston, MA \\
nfulton@mit.edu}
\and
\IEEEauthorblockN{Sanjay Krishna Gouda}
\IEEEauthorblockA{\textit{AWS AI Labs} \\
New York, NY \\
skgouda@amazon.com}
\and
\IEEEauthorblockN{Haifeng Qian}
\IEEEauthorblockA{\textit{AWS AI Labs} \\
New York, NY \\
qianhf@amazon.com}
}

\maketitle

\begin{abstract}
Large Language Models are powerful tools for program synthesis and advanced auto-completion, but come with no guarantee that their output code is syntactically correct.
This paper contributes an incremental parser that allows early rejection of syntactically incorrect code, as well as efficient detection of complete programs for fill-in-the-middle (FIM) tasks.
We extend the Earley parsing algorithm to allow for left and right quotients of context-free grammars, and develop methods to handle quotienting of several context-sensitive features present in the grammars of many common programming languages.
The result of these contributions is an efficient, general, and well-grounded method for left and right quotient parsing.

To validate our theoretical contributions---and the effectiveness of certain design decisions---we evaluate our method on the particularly difficult case of FIM completion for Python~3, with syntax-correctness constraints. Our results demonstrate that constrained generation can significantly reduce the incidence of syntax errors in recommended code. %
\end{abstract}


\section{Introduction}

In recent years, several AI-assisted code generation tools have been released, such as Amazon Q Developer \cite{amazonwebservicesinc.AICodeGenerator2023} and GitHub Copilot \cite{githubinc.GitHubCopilotYour2023}. 
These models take advantage of powerful Large Language Models (LLMs) fine-tuned on code, such as Starcoder \cite{liStarCoderMaySource2023}, OpenAI Codex \cite{chenEvaluatingLargeLanguage2021}, Deepseek Coder \cite{deepseekai2024deepseekcoderv2breakingbarrierclosedsource}, Code Llama \cite{roziere2024codellamaopenfoundation}, and Codestral \cite{Codestral},
and they often perform well on code generation tasks. 

These code generation systems are typically presented as an IDE plugin.
When the plugin is activated, the current cursor position, termed the \emph{insertion point}, splits the existing code into a \emph{left context} and \emph{right context}.
The contexts are \emph{tokenized} to obtain a sequence of LLM tokens---sequences of one or more text characters; i.e. the right parentheses character, or the keyword ``\texttt{def}''.\footnote{Note well: Throughout this paper, we use ``token" to mean an LLM token, ``symbol" to mean the atomic unit of a lexer's output, and ``lexeme'' to mean a category of symbol. For example: \texttt{foo} and \texttt{23} are symbols, the category of all identifiers or all integers are lexemes, and \textsc{eos} is a token. We use ``element'' to refer to the terminals and nonterminals of a CFG.}

LLMs are auto-regressive predictors.
Given $n$ LLM tokens $(s_1, \ldots, s_n)$, the model calculates the probabilities of a token that follows $p(s_{n+1}|s_1, \ldots, s_n)$.
A single token $s_{n+1}$ is chosen (via sampling, greedy selection, beam search, etc.), and is appended to the input sequence for the next iteration.
This is repeated until the end of sequence token (\textsc{eos}) is generated.

\label{section:fim-special-tokens}

For code completion that involves the right context, known as the \emph{fill-in-the-middle} (FIM) task, it is still possible to use this procedure by adding several ``control'' tokens to the LLM's vocabulary: \textsc{FIM-Prefix}, \textsc{FIM-Suffix}, and \textsc{FIM-Middle} \cite{bavarianEfficientTrainingLanguage2022}.
Files from the dataset are randomly cut into three sections: ``prefix'', ``middle'', and ``suffix''.
The sections are arranged into a sequence of the form \textsc{FIM-Prefix} $\circ$ \textsc{Enc}(prefix) $\circ$ \textsc{FIM-Suffix} $\circ$ \textsc{Enc}(suffix) $\circ$ \textsc{FIM-Middle} $\circ$ \textsc{Enc}(middle) $\circ$ \textsc{eos}, where $\textsc{Enc}$ represents the tokenization process, and $\circ$ is concatenation.
The LLM is trained on such sequences.
During inference, if the input is the above sequence up to and including \textsc{FIM-Middle}, the highest probability sequence of outputs should be \textsc{Enc}(middle) $\circ$ \textsc{eos}.

Language generation models are not perfect.
They may generate \texttt{EOS} too early or too late,
and are often brittle with respect to minor details such as the number of spaces.
Because they treat code as unstructured text, rather than as a member of a programming language with a formal grammar, they may generate incorrect syntax altogether.
One solution is to incorporate the language's inherent structure. 

Constrained generation has been used in code completion \cite{jonesLlamaAddGrammarbased2023, slattonAddedContextFree2023,willardEfficientGuidedGeneration2023,takerngsaksiriSyntaxAwareOntheFlyCode2023} and structured data extraction tasks \cite{microsoftTypeChat2023,sengottuveluJsonformerBulletproofWay2023,automorphicTrex2023, microsoftGuidance2023,sriLQML2023,willard2023efficient} to incorporate a language's structure. %
Constrained generation can also be used to improve alignment of tokens for generations that begin in the middle of a word \cite{athiwaratkun2024tokenalignmentcharactermatching}.

The common idea behind all of these methods is to test potential tokens as the LLM generates them, and select a different one if a given token would cause the generation not to be a prefix of a valid program; or equivalently, create a set of legal tokens, and sample only tokens in this set.

However, existing methods do not allow for FIM generation.
In addition to ensuring that the generated code is a prefix of a valid program, FIM generation requires that the generated code is able to ``connect'' to the right context---a tricky prospect when neither the left nor the right contexts are valid programs in isolation.
Additionally, these methods often do not account for the complexities of real-world programming languages, which usually exhibit context sensitive lexing behavior.

Several methods focus on AST node prediction \cite{mukherjeeNeuralProgramGeneration2021,liCodeCompletionNeural2018,kimCodePredictionFeeding2021,liuUnifiedMultitaskLearning2022,svyatkovskiyPythiaAIassistedCode2019} as a way to generate code, potentially filling in the middle; however, these methods require an AST with holes to be readily constructed.
Such an AST may not be available, for example, if the left or right contexts are cut off in the middle of a symbol, as is common in code completion settings.
Even in cases where it is possible to build an AST with holes, node insertion does not allow for the full space of edits possible with text insertion: new characters may cause modifications to ancestor or sibling AST nodes.
For example, in Python, the transformation from ``\texttt{\{a:$\square$\}}'', where $\square$ represents an AST hole at the insertion point, to ``\texttt{\{a:b for a,b in c\}}'' requires changing an ancestor node from a dictionary literal to a dictionary comprehension.

This paper presents a method that is able to perform constrained FIM generation for a real programming language, even with these requirements.
We first contribute an extension to the Earley algorithm that allows it to compute the right quotient of a context-free language, given a suffix.
We then detail several features of real programming languages that are not context-free, and present modifications to the quotienting operation and incremental parser to handle these context-sensitive features.
We integrate all of these methods into a proof-of-concept parser for Python 3, and use it for constrained generation with a LLM.
Through a series of experiments, we show that this system is often able to generate valid code in cases where unconstrained generation fails.

\begin{figure}
    \centering
    \includegraphics[width=0.6\linewidth,page=4]{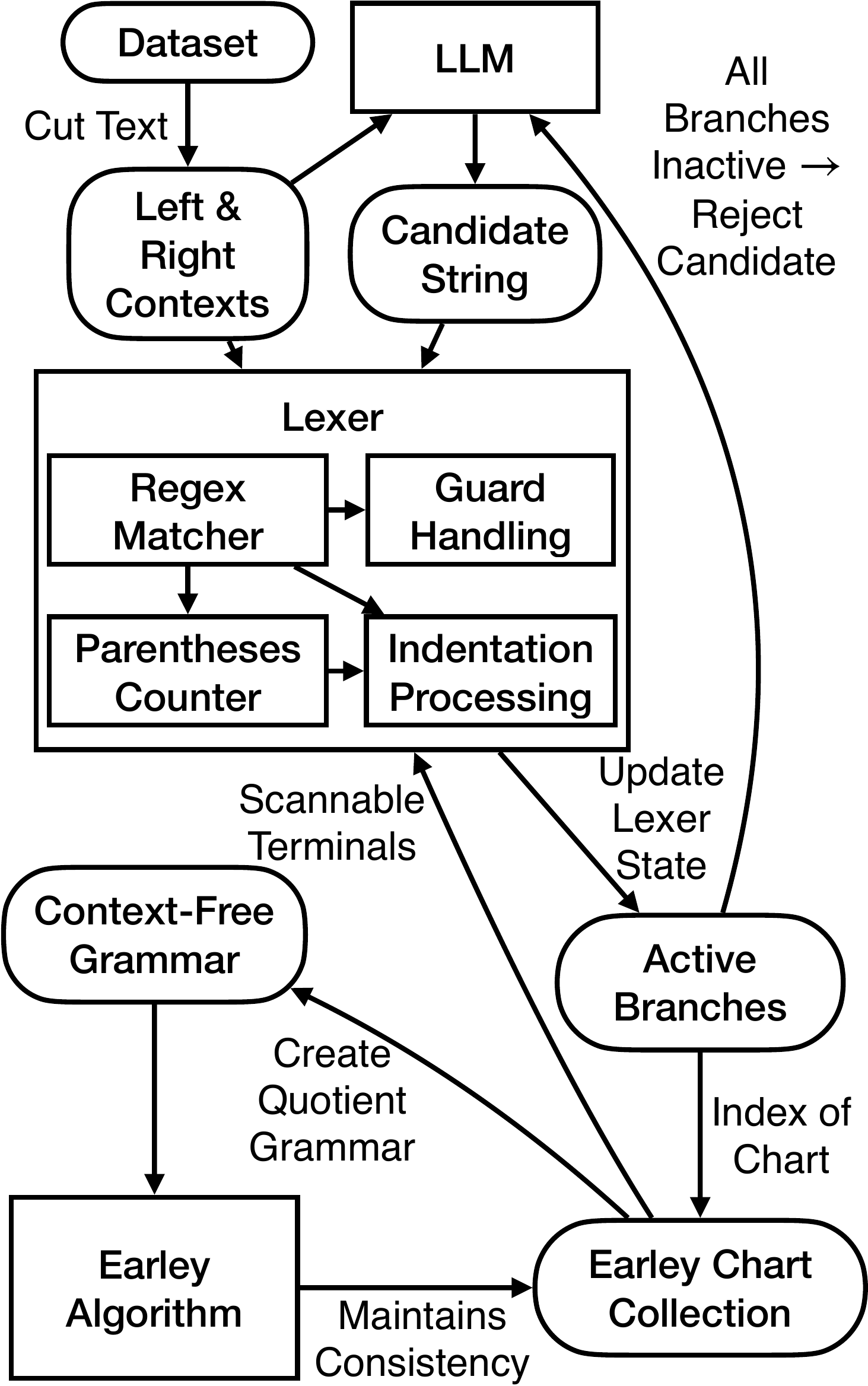}
    \caption{Conceptual overview of our method, integrated within a constrained generation process. Rounded rectangles represent data; cornered rectangles represent processes.}
    \label{fig:conceptual-overview}
\end{figure}

\section{Background}
\label{sec:background}

\subsection{Automata and Regular Languages}

A Nondeterministic Finite Automaton (NFA) is a tuple $\nfaTuple$, where $\nfaStateSpace$ is a set of states, $\nfaAlphabet$ is an alphabet, $\nfaInitialState \in \nfaStateSpace$ is an initial state, $\nfaTransition : \nfaStateSpace \times \nfaAlphabet \rightarrow \powerset{\nfaStateSpace}$ is a transition function, and $\nfaFinalStates \in \powerset{\nfaStateSpace}$ is a set of final states.
A string $\alpha = (\alpha_1, \ldots, \alpha_n) \in \nfaAlphabet^{n}$ is \emph{accepted} by a NFA if $\exists (q_0, \ldots, q_n), q_0 = \nfaInitialState \land \forall i \in [1..n], q_i \in \nfaTransition(q_{i-1}, \alpha_i) \land q_n \in \nfaFinalStates$.
Language $\reg$ is \emph{Regular} iff there exists a NFA $\phi$ that accepts exactly the strings which are members of that language; $\reg$ is said to \emph{correspond} to $\phi$. 

\subsection{Context-Free Grammars and Languages}

A context-free grammar (CFG) is a tuple $\cfgTuple$ where $\cfgNonterminals$ is the set of nonterminal elements, $\cfgAlphabet$ is an alphabet of terminal elements, $\cfgRules$ is a set of production rules, each of type $\cfgNonterminals \times (\cfgNonterminals \cup \cfgAlphabet)^*$ representing a nonterminal element and a possible expansion, and $\cfgStart$ is a top-level rule.

A CFG represents all strings that can be obtained by starting with $\cfgStart$ and recursively expanding all nonterminal elements with some rule in $\cfgRules$ until only terminal elements remain.
A language is context-free iff there exists a CFG that represents all strings in that language; we use $\langOfCFG{G}$ to represent the language described by CFG $G$.

\section{Problem Statement}
\label{sec:problem-statement}

\label{subsec:incremental-parsing-desirata}

When the LLM proposes a token, the parser must decide whether this token is acceptable. An \texttt{EOS} token is acceptable if the text that has been generated so far is a complete program; i.e. the parser performs a \emph{membership query}. Any other token is acceptable if the resulting program is completable; i.e. there is some text which can be generated such that the full program would be complete.

\begin{definition}
Let $\mathcal{L}$ be a language over alphabet $\Sigma_{\mathcal{L}}$.
An incremental recognizer for $\mathcal{L}$ is a function of type $ \Sigma_{\mathcal{L}}^* \rightarrow \mathbb{B}$ that determines for string $\alpha$ whether $\exists \gamma \in \Sigma_{\mathcal{L}}^*, \alpha \circ \gamma \in \mathcal{L}$.
\end{definition}

The \emph{right quotient} is a binary operator over two languages.
For languages $\mathcal{L}, \mathcal{L}_R$ over $\charAlphabet$, the right quotient $\mathcal{L} / \mathcal{L}_R = \{l \in \charAlphabet^* | \exists r \in \mathcal{L}_R, l \circ r \in \mathcal{L}\}$ \cite{linzIntroductionFormalLanguages2012}.
Note that if $\mathcal{L}_R$ only contains string $\beta$ (denoted as $L(\beta)$), the right quotient $\mathcal{L} / L(\beta) = \{l \in \charAlphabet^* | l \circ \beta \in \mathcal{L}\}$.
The \emph{left quotient} is defined analogously.

Our desired functionality can be captured as follows:
\begin{problem}
    Let $\mathcal{L}$ be a language over alphabet $\Sigma_{\mathcal{L}}$.
    For right context $\beta \in \Sigma_{\mathcal{L}}^*$, construct an incremental recognizer, and a membership query function, for $\mathcal{L} / L(\beta)$.
\end{problem}

Given a left context $\alpha$ and a right context $\beta$, when an \texttt{EOS} token is proposed, we can use language membership queries to check if $\alpha \in \mathcal{L} / L(\beta)$.
If character $c$ is proposed, we can use the incremental recognizer to test $\exists \gamma, \alpha \circ c \circ \gamma \in \mathcal{L} / L(\beta)$.

Finally, we note that these operations should be efficient; in particular, if the incremental recognizer or membership query functions are called for several strings with a common prefix and right context, it should be possible to reuse most of the work, rather than parsing the whole input every time---ideally resulting in a constant-time overhead per generated token; linear time in generation length.

\subsection{Why Not Just Run The Parser?}
\label{subsection:checked-unconstrained-why-not}

There is a simple way to ``incrementally'' test for quotient language membership:
for every character created during constrained generation, concatenate the left context, the generated code up to that character, and the right context; then, just invoke the language's parser using this string.

However, this approach has two main drawbacks.
First, it cannot test for incremental parsability, especially relative to a quotient language---the utility of the parser with unconstrained generation is limited to checking the validity of a \texttt{EOS} token.
Therefore, this approach, which we refer to as \emph{checked unconstrained generation}, can not be used for early rejection. It allows for the generation of code with no possible completion.

Second, checked unconstrained generation requires parsing the entire input string for each character, contradicting our efficiency requirements.
As we show in our experiments, this results in unfavorable asymptotic complexity: linear-time per generated token; quadratic time in generation length.

\section{Earley Parsing}

We first provide a short overview of the well-known Earley algorithm \cite{earleyEfficientContextfreeParsing1970} for parsing context-free languages.
We assume that a CFG $G = \cfgTuple$ is given, and describe a modification to the Earley algorithm that allows for the calculation of a CFG corresponding to $L(G) / \mathcal{R}$ for regular language $\mathcal{R}$.

\label{section:earley_parsing}

Let $X = (X_0, \ldots, X_{n-1})$ be an input string; we want to test if $X \in L(G)$.
The state of the Earley algorithm is a sequence of \emph{charts} $c_0 \ldots c_n$. 
Each chart contains a set of Earley \emph{items}.
Each item is of the form $[E \rightarrow \alpha \bullet \beta \: (i)]$, where:
\begin{itemize}
    \item $E \in \cfgNonterminals$ is a nonterminal,
    \item $\alpha$ and $\beta$ are strings of \emph{elements} ($\cfgAlphabet \cup \cfgNonterminals$), and
    \item $i$ is a \emph{span start} index (referencing $c_i$).
\end{itemize}

Chart $c_0$ is initialized with item $[\cfgStart \rightarrow \bullet \alpha \: (0)]$ for all production rules $(\cfgStart, \alpha) \in \cfgRules$.
Upon being added to chart $c_i$, each item is placed in a queue to be processed.

If an added item is of the form $[E \rightarrow \alpha \bullet a \beta \: (s)]$, for terminal $a$, it is processed by the \textit{scanner}.
If $X_i = a$, item $[E \rightarrow \alpha a \bullet \beta \: (s)]$ is added to chart $c_{i+1}$.
Note that $a$ is termed a \emph{scannable terminal} for chart $c_i$.
The Earley item is not added if chart $c_{i+1}$ already contains a duplicate, otherwise, the new item is also placed on the processing queue.

If the added item is of the form $[E \rightarrow \alpha \bullet A \beta \: (s)]$, the \emph{predictor} adds items $[A \rightarrow \bullet \gamma \: (i)]$ to chart $c_i$ for every BNF production $(A, \gamma) \in \cfgRules$.
As each chart is a set of items, duplicate items are not added if $A$ is predicted twice.

Finally, items of the form $[E \rightarrow \alpha \bullet \: (s)]$ are processed by the \emph{completer}.
The algorithm searches chart $c_s$ for \emph{completable} items of the form $[X \rightarrow \gamma \bullet E \delta \: (n)]$; for each of these, an item of the form $[X \rightarrow \gamma E \bullet \delta \: (n)]$ is added to chart $c_i$.

$X$ is \emph{recognized} by an Earley parser iff chart $c_n$ contains any complete items, where the item's grammar element is $\cfgStart$.
Regardless of this, if $c_n$ has any scannable terminals $T \subseteq \Sigma$, $X$ is a prefix of a member of $L(G)$, and $\forall t \in T, (X_0, \ldots, X_{n-1}, t)$ is also a prefix of some member of $L(G)$. 

\subsection{Earley Algorithm Applied to NFAs}

Traditionally, the Earley algorithm is viewed as having a sequence of charts, where each chart is partial parse state ``in between'' characters of $X$.
We instead represent the algorithm state as the nondeterministic transition system $(\nfaStateSpace = \{c_0, \ldots, c_n\}, \nfaAlphabet, \nfaInitialState=c_0, \nfaTransition, \nfaFinalStates = \{c_n\})$, where
\begin{equation*}
    \nfaTransition(c_i, x) = \begin{cases}
    x = X_i & \{c_{i+1}\} \\
    \text{otherwise} & \emptyset
\end{cases}
\end{equation*}

We note that this transition system is identical to the NFA corresponding to the language containing exactly the string $X$, with a chart for each NFA state.
After adding the initial items to $c_0$, the scanner, predictor, and completer are iteratively run until the collection of charts reaches a fixpoint.
Rather than always scanning into the next chart sequentially, the scanner ``follows'' the arrows of the transition system.
The predictor and completer are unmodified; however, some additional bookkeeping is required to ensure that newly added completable items are always processed by the completer.

It is possible to use more complex transition systems, corresponding to arbitrary regular languages $\mathcal{R}$ rather than single strings; if so, this algorithm determines if $L(G) \cap \mathcal{R} \neq \emptyset$---effectively ``parsing'' a regular language with the CFG\footnote{Even when only parsing strings, this approach is useful for representing the Earley algorithm state by only an index, enabling cheap copies.}
An example of this view is illustrated in the top half of Figure \ref{fig:earley-quotient}.

\subsection{Quotient Extraction}

\begin{figure}
    \centering
    \includegraphics[width=0.94\linewidth,page=1]{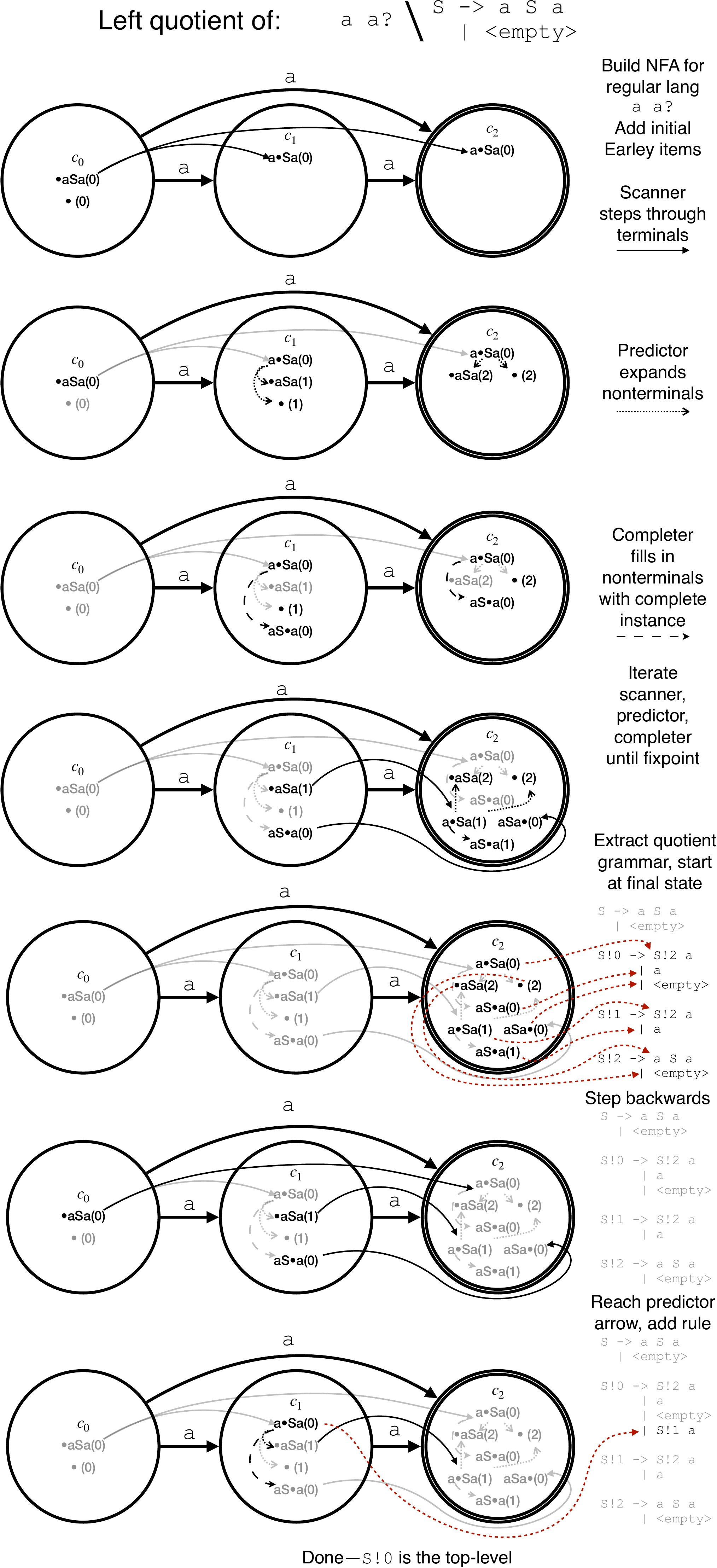}
    \caption{We ``parse'' a regular language with a CFG, using the Earley algorithm modified to work over NFAs. The algorithm tracks how each Earley item is created by the scanner, predictor, and completer, shown as each Earley item are tracked, shown as thin solid, dotted, and dashed arrows respectively. When constructing the quotient grammar, each item in the final Earley chart corresponds to a production rule, representing the sequence necessary in order to complete each item. The algorithm finds all prediction arrows on the path from the initial state to the final state, and adds rules corresponding to each of those. Note that Earley items omit the ``S $\rightarrow$'', as there is only one nonterminal in this example.}
    \label{fig:earley-quotient}
\end{figure}

If we add additional bookkeeping to track how each Earley item is created, these charts contain an implicit representation of the left quotient $\mathcal{R} \setminus L(G)$.
The algorithm presented in Figure \ref{algorithm:extract-quotient-lang} extracts this as an explicit CFG; the extraction procedure is illustrated in the bottom half of Figure \ref{fig:earley-quotient}.
Note that while duplicate Earley items are still discarded after creation, the creation method of the duplicate item must be added to the original item for the quotient extraction algorithm to return the correct result.

Finally, we can obtain the right quotient $L(G) / \mathcal{R}$:
Let \textsc{Reverse} reverse a language, such that string $X \in \mathcal{L}$ iff the reverse string $\overleftarrow{X} \in \textsc{Reverse}(\mathcal{L})$. 
The CFL corresponding to a CFG may be reversed by reversing each production; the regular language corresponding to a NFA may be reversed by inverting transitions and swapping the initial and final states.
It holds that $L(G) / \mathcal{R} = \textsc{Reverse}(\textsc{Reverse}(\mathcal{R}) \setminus \textsc{Reverse}(L(G)))$.

\subsubsection*{Example}

Consider the incomplete Python function call ``\texttt{foo(a,}''.
The lexer emits four symbols: \texttt{ID LPAR ID COMMA}.
Using a simplified, representative subset of Python's grammar, there exist rules like the following: %

\begin{verbatim}
Expr -> ID | Call
Call -> Expr LPAR Args RPAR
Args -> <empty> | Expr ArgCont
ArgCont -> <empty> | COMMA Expr ArgCont
\end{verbatim}

Based on this grammar, an \texttt{Expr} and a \texttt{Call} are both started at index 0 (before the first \texttt{ID}), an \texttt{Args} is started at index 2, and an \texttt{ArgCont} started at index 3.
There are also additional nonterminals that are not ``open'' at the end, such as the \texttt{Expr} containing only the \texttt{ID} between index 2 and 3.
The core of our algorithm is to find the completion of each open nonterminal.
For example, chart $c_4$ contains the item $[\texttt{ArgCont} \rightarrow \texttt{COMMA} \bullet \texttt{Expr ArgCont} (3)]$, meaning that the the \texttt{ArgCont} that began at index 3 has ``seen'' a \texttt{COMMA}, and it can be completed by a \texttt{Expr} and another \texttt{ArgCont}:
\begin{verbatim}
ArgCont!3 -> Expr ArgCont
\end{verbatim}

The algorithm works backwards through item creation methods to find $[\texttt{Args} \rightarrow \texttt{Expr} \bullet \texttt{ArgCont} (2)]$ in chart $c_3$, indicating that the \texttt{Args} that began at index 2 can be completed by finishing the \texttt{ArgCont} that began at index 3:
\begin{verbatim}
Args!2 -> ArgCont!3
\end{verbatim}

Similarly, there is an item $[\texttt{Call} \rightarrow \texttt{Expr LPAR} \bullet \texttt{Expr RPAR} (0)]$ in chart $c_2$, indicating that the \texttt{Call} that began at index 0 is completed after the \texttt{Args} beginning at index 2 is completed, and then a \texttt{RPAR} is generated:

\begin{verbatim}
Call!0 -> Args!2 RPAR
\end{verbatim}

The algorithm then encounters $[\texttt{Expr} \rightarrow \bullet \texttt{Call} (0)]$ in chart $c_0$, indicating that the \texttt{Expr} that began at index 0 is completed after the \texttt{Call} that began at index 0 is done:

\begin{verbatim}
Expr!0 -> Call!0
\end{verbatim}

The algorithm also notes that this Earley item was created as part of the initialization phase, and adds \texttt{Expr!0} as a top-level nonterminal in the quotient grammar.

Finally, the algorithm encounters the item $[\texttt{Call} \rightarrow \bullet \texttt{Expr LPAR Expr RPAR} (0)]$ in chart $c_0$. 
This corresponds to the case where \texttt{foo} returns a function, which itself is called, and so the algorithm adds a second case to \texttt{Call!0}:

\begin{verbatim}
Call!0 -> Args!2 RPAR 
        | Expr!0 LPAR ARGS RPAR
\end{verbatim}

\begin{figure}
    \begin{algorithmic}[1]
        \Procedure{ExtractGrammar}{$G = \cfgTuple, \reg = \nfaTuple, \{c_q | q \in \nfaStateSpace\}$}
        \State Frontier $\leftarrow \{(s, q) | s \in c_q, q \in \nfaFinalStates\}$
        \State $N \leftarrow \{(E_i, \beta) | ([E \rightarrow \alpha \bullet \beta \: (i)], q) \in \text{Frontier}\}$ \label{line:rule-initialization}
        \While{(s, q) = \textsc{Pop}(Frontier)}
            \State \textbf{for} creation method $c$ of $s$ \textbf{match} s, c
            \Indent
            \State $*$, \emph{scanned} from item $s'$ in $c_j$
            \Indent
            \State \textsc{PushOnce}(Frontier, $(s', j)$) \label{line:scanner-frontier}
            \EndIndent
        \State $*$, \emph{predicted} from $s' = [F \rightarrow \alpha \bullet E \beta (i)]$ in $c_q$
        \Indent
            \State $N \leftarrow N \cup \{(F_i, E_q \circ \beta)\}$\label{line:predictor-rule}
            \State \textsc{PushOnce}(Frontier, $(s', q)$) 
        \EndIndent
        \State $*$, \emph{completed} into item $s'$ in $c_j$
        \Indent
            \State \textsc{PushOnce}(Frontier, $(s', j)$)
        \EndIndent
        \State $[\cfgStart \rightarrow \bullet \alpha \: (\nfaInitialState)]$, \emph{initialized} in $c_{\nfaInitialState}$
        \Indent
            \State $N \leftarrow N \cup \{(\cfgStart', \cfgStart_q)\}$ \label{line:initialized-rule}
        \EndIndent
        \EndIndent
            \State \textbf{end for match}
        \EndWhile
        \State \Return $(\cfgNonterminals \cup \{\cfgStart'\}, \Sigma, N \cup \cfgRules, \cfgStart')$
           \EndProcedure
    \end{algorithmic}
    \caption{Extract quotient language from CFG by traversing Earley item creation methods.
    For each chart $c_i$, and nonterminal $g \in \cfgNonterminals$, this algorithm computes the grammar representing how to complete any Earley items for $g$ that originate in $c_i$.}
    \label{algorithm:extract-quotient-lang}
\end{figure}

\section{Lexed Context-Free Languages}

\label{section:lcfls}

Most programming languages are not context-free;
for example, nearly every programming language uses a lexer to transform source text into a sequence of language symbols.
A programming language specifies a number of lexemes; each lexeme, in turn, is recognized by a regular language.
Furthermore, most programming languages treat whitespace and comments in a special manner.
We must take both of these factors into account.

The lexing procedure is often described as following the leftmost-longest, or ``maximal munch'', lexing rule.\footnote{\label{footnote:parsing-weird-behavior}Many programming languages, sometimes contradicting their documentation, do not actually use leftmost-longest lexing. This effect can be seen in Python with the string \texttt{0or 1}. Rather than lexing as the three symbols \texttt{0}, \texttt{or}, and \texttt{1}, Python's lexer raises an error because it first ``commits'' to reading \texttt{0o} as an octal number. Java has a similar (albeit officially documented) effect where \texttt{non-sealedclass} is not interpreted as the two keywords \texttt{non-sealed} and \texttt{class}. This illustrates an important idea: many real-world programming languages are not formally specified, and it is unlikely that any re-implementation will capture all of a language's properties. While our implementation does capture this behavior, our ultimate goal must be to empirically improve the quality of code generation by a meaningful amount, not to create a perfect right-quotienting incremental parser.}
This rule states that the longest prefix of the source string that matches any lexeme's language must be accepted as the first symbol; the lexer will then be run on the remainder of the source string.
The procedure repeats until there is no text remaining.

\begin{figure}
    \centering
    \includegraphics[width=\linewidth,page=3]{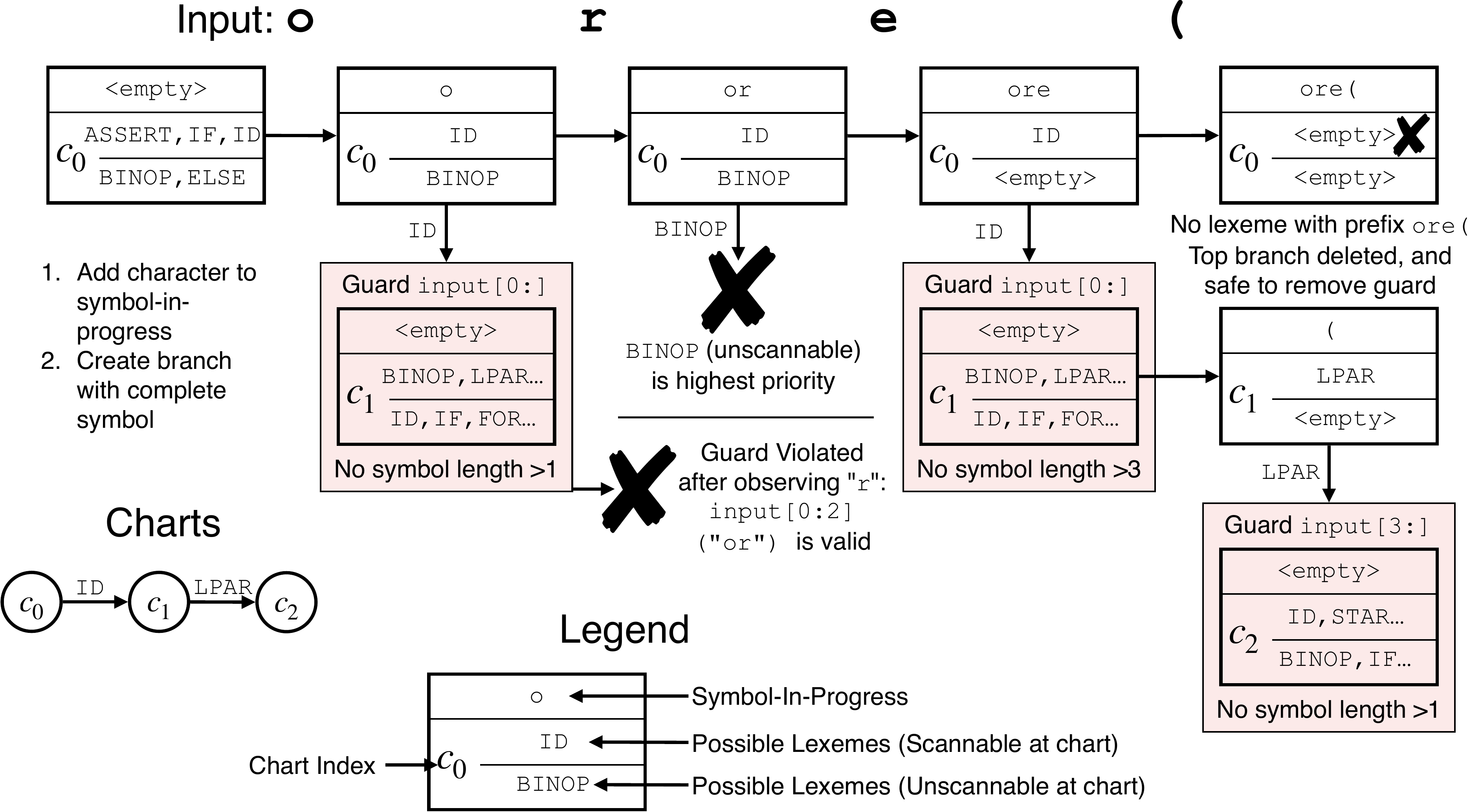}
    \caption{Example of lexer actions, and interaction with parser, for the string ``\texttt{ore(}''. Each box represents a lexer state. }
    \label{fig:lex-parse-example}
\end{figure}

\subsection{Incremental Lexing}

There are additional complications that arise when lexing in an incremental setting.
Most importantly, an incremental lexer cannot access any lookahead.
When a lexeme is matched, the lexer cannot determine, in the moment, if the symbol-in-progress should be part of its output stream, or if a longer symbol would be appropriate with additional text.

Our solution to this predicament is to create a branching lexer---it tries both options, then discards the branch that does not reflect the text actually received.
Given symbol-in-progress $sip$, and a received character $x_i$, the first branch reflects a world in which $x_i$ is \emph{not} the final character of the symbol.
The symbol-in-progress is extended to include $x_i$.
This branch may be deleted upon a lex failure; i.e. when the symbol-in-progress no longer is a prefix of any lexeme.

The second branch reflects the case where $x_i$ \emph{is} the final character of a symbol; i.e., $sip + x_i$ would be matched by the leftmost-longest rule.
The highest-priority lexeme that matches $sip + x_i$ is fed to the parser.
To protect against the case in which $sip + x_i$ is \emph{not} the longest match after additional text is given, the branch contains a ``branch guard''.
If a longer sequence of characters matches some lexeme starting at the same input index, the branch guard causes the branch to be deleted.

Note that this structure prevents an exponential explosion of branches; the same condition that triggers a further branching of the first branch causes the deletion of the second branch\footnote{Note that leftmost-longest matching allows for further branching of the second branch, without deleting the first branch. However, this almost never occurs in practice, and is always resolved within a few characters---the necessary conditions are the same as those which cause Python's undocumented behavior mentioned in Footnote \ref{footnote:parsing-weird-behavior}. Our implementation of Python's observed lexing behavior deletes the first branch if it does not immediately further branch on the next character, guaranteeing a limit of 2 branches.}.

\begin{figure}
     \begin{algorithmic}[1]
        \Procedure{IncLex}{$G = \cfgTuple, (X_0, \ldots, X_{n-1})$}
\State $c_0 \leftarrow \textsc{InitEarley}(G)$
\State $B \leftarrow \{(c_0, \text{`'}, ST(c_0), \Sigma \setminus ST(c_0), \{\})\}$\Comment{Branches}
\For{$x_i \in (X_0, \ldots, X_{n-1})$}
    \State $B' \leftarrow \{\}$
    \For{$b = (c_b, sip, st, st_{\text{unsafe}}, \textit{guards}) \in B$}
      \State $\textit{guards}' \leftarrow$ Advance \textit{guards} with $x_i$
      \State \textbf{if} $\exists g \in \textit{guards}'$ violated \textbf{continue end if} 
      \State $sip' \leftarrow sip + x_i$
      \State $st' \leftarrow \{t \in st | sip' \text{ is prefix of } t\}$ \label{line:sip-prefix-of}
      \State $st_\text{unsafe}' \leftarrow \{t \in st_\text{unsafe} | sip' \text{ is prefix of } t\}$
      \State \textbf{if} $st'$ subsumed by $st'_\text{unsafe}$ \textbf{continue end if} \label{line:subsumed_check}
      \State Add $(c_b, sip', st', st_\text{unsafe}', \textit{guards}')$ to $B'$  
      \State $t_{\text{select}} \leftarrow$ Highest priority terminal matching $sip$
      \State $c^* \leftarrow \textsc{EarleyStep}(c_b, t_\text{select})$
      \State $\textit{guards}^* \leftarrow \textit{guards}' \cup \{sip', |sip'|\}$
      \State Add $(c^*, \text{`'}, ST(c^*), \Sigma \setminus ST(c^*), \textit{guards}^*)$ to $B'$
    \EndFor
    \State $B \leftarrow B'$
\EndFor
        \EndProcedure
    \end{algorithmic}
    \caption{Incremental branch-based lexer and interaction with parser. $ST(c)$ returns the scannable terminals of chart $c$.}
    \label{algorithm:lexer-parser-interaction}
\end{figure}

The algorithm that implements this is further described in Figure \ref{algorithm:lexer-parser-interaction}, and a short example is illustrated in Figure \ref{fig:lex-parse-example}.

We would like to draw attention to the subroutine of the algorithm that determines when symbol-in-progress $sip$ is not the prefix of any legal lexeme (lines \ref{line:sip-prefix-of}-\ref{line:subsumed_check}).
In addition to ensuring that there exists some suffix $\alpha$ such that $sip + \alpha$ matches a scannable terminal, the lexer must also ensure that $sip + \alpha$ does not match a lexeme of higher priority that is not a scannable terminal. 
Due to Python's specific structure, we did not encounter any cases where every possible completion of the symbol-in-progress would be matched by a higher-priority non-scannable lexeme, except in cases where no completions match a scannable terminal at all.
However, in languages with a more exotic priority structure, this check would likely require a regular language subset test.

Finally, we note that comments and whitespace can be handled by always allowing the lexemes for these as scannable terminals, when any scannable terminals exist. 
When matched, the lexer simply does not pass the whitespace to the parser.

\subsection{Right Quotient of Lexed Languages}

\begin{figure}
    \centering
\begin{minted}[escapeinside=||]{text}
|\mintedComment{Clean Boundary: Index 0}|
x = |\highlightCode{"#'#"#"#\textbackslash{}n} |
|\mintedComment{Unclosed Double Quotes: Index 1}|
"foo|\highlightCodeBlue{"}\highlightCode{#'#"#"#\textbackslash{}n} |
|\mintedComment{Unclosed Long Quotes: Index 1}|
"""foo""|\highlightCodeBlue{"}\highlightCode{#'#"#"#\textbackslash{}n} |
|\mintedComment{Unclosed Single Quotes: Index 3}|
'foo|\highlightCodeBlue{"#'}\highlightCode{#"#"#\textbackslash{}n} |
|\mintedComment{Unclosed Quotes with Backslash: Index 5}|
"foo\|\highlightCodeBlue{"#'#"}\highlightCode{#"#\textbackslash{}n} |
|\mintedComment{Comment: Index 8}|
#foo|\highlightCodeBlue{"#'#"#"#}\highlightCode{\textbackslash{}n}|
|\mintedComment{Unclosed Triple Quotes: Error}|
"""foo|\highlightCodeRed{"#'#"#"#\textbackslash{}n}|
\end{minted}
    \cprotect\caption{When the right context starts with \verb|"#'#"#"#\n|, there are many possibilities for how much of it is captured by the last symbol in the quotient text. While this example is artificial, most natural right contexts still have several possible indices.}
    \label{fig:capture_diff_amount_of_right_context}
\end{figure}

\label{section:lcfl_right_quotient}

The presence of a lexer adds complexity to the right quotient operation.
An immediate issue is that the beginning of the right context might be the continuation of a symbol from the quotient language. 
Consider the Python program shown in Figure \ref{fig:capture_diff_amount_of_right_context}. %
There are many possible symbol boundaries in the right context, depending on the quotient text.
However, it is unknown which case is true ahead of time, as the quotient text is yet to be generated!

Our approach to this issue uses the same branching structure introduced in our lexer.
We first ``guess'' how many characters at the beginning of the right context are a continuation of the last symbol of the quotient text, and then compute the potential symbols that may result.
The algorithm presented in Figure \ref{algorithm:calc_boundaries_of_right_context} efficiently calculates all indices for which it is possible for a symbol continued from the quotient text to end, providing a result similar to that shown in Figure \ref{fig:capture_diff_amount_of_right_context}.
This algorithm is further explained and illustrated in Figure \ref{fig:lex-right-quotient}.
A branch is created for each index, and the parser is run for each branch, for all text after that branch's index.

For example, one of the lexer's branches begins at index 1, corresponding to both the second and third examples from Figure \ref{fig:capture_diff_amount_of_right_context}.
Within this branch, the lexer begins lexing from the second character---beginning with a comment, then a newline---and the algorithm builds a quotient language from the emitted symbols using the method described in Section \ref{section:earley_parsing}.
This is repeated for all branches.

Additionally, the quotient language for this branch must reflect that there is still a double-quote character unaccounted for in the right context.
The only lexemes which may include this double quote (and not capture any parts of the right context past the first character, as these were already lexed) are those related to strings; we therefore restrict the quotient grammar such that every member of the language must end in a string.
Finally, the lexer for the quotient language ensures that the final symbol captures exactly the first character; i.e. it does not accept a quotient text ending in a backslash, as that would capture beyond the first character of the right context.

In our experiments---described later---this procedure results in a median of 5 unique starting indices, and a maximum of 10, including index 0 (no continuation).

\begin{figure}
\begin{algorithmic}[1]
\Procedure{CalcRQIdx}{Right Context = $(S_1, \ldots, S_n)$}
\State $\forall g \in \Sigma, \lexemeNFA{g} = (\nfaStateSpace_g, \nfaAlphabet_\mathcal{L}, \nfaInitialState^g, \nfaTransition_g, \nfaFinalStates_g) \leftarrow \textsc{Matcher}(g)$
\State $\forall g \in \Sigma, Q_g^0 \leftarrow \bigcup_{x \in \charAlphabet} \nfaTransition_g(\textsc{Reachable}(\nfaInitialState^g), x)$ \label{line:cond-3-start}
\For{$S_i \in (S_1, \ldots, S_n)$}
\State $\forall g \in \Sigma, Q_g^i \leftarrow \nfaTransition_g(Q_g^{i-1}, S_i)$
\EndFor \label{line:cond-3-end}
\For{$g \in \Sigma$} \label{line:for_g}
    \For{$S_i \in \textsc{Reverse}(S_1, \ldots, S_n)$} \label{line:for_idx_char_rev_suffix}
    \Comment{Constraint: $\forall i, j \in [1\ldots n], q \in \nfaFinalStates_g^{i\rightarrow j},  \nfaTransition_g(q, (S_{i+1},\ldots, S_j)) \cap \nfaFinalStates_g \neq \emptyset$}
        \State $\nfaFinalStates_g^{i \rightarrow i} \leftarrow \nfaStateSpace^i_g \cap \nfaFinalStates_g$
        \State $\forall j \in [i\ldots n], \nfaFinalStates_g^{i-1 \rightarrow j} \leftarrow \nfaTransition_g^{-1}(\nfaFinalStates_g^{i \rightarrow j}, S_i)$ \label{line:step_back_future_ends}
        \State $\forall j \in [i-1 \ldots n]$ 
        \For{$j \in [n \ldots i-1]$} \label{line:remove-subsumed}\Comment{Remove Subsumed}
            \State $\nfaFinalStates_g^{i-1 \rightarrow j} \leftarrow \nfaFinalStates_g^{i-1 \rightarrow j} \setminus \bigcup_{k > j} \nfaFinalStates_g^{i-1 \rightarrow k}$
        \EndFor
    \EndFor \label{line:end_for_idx_char_rev_suffix}
\EndFor
\State \Return $\{j | g \in \Sigma, j \in [1\ldots n] | \nfaFinalStates_g^{0 \rightarrow j} \neq \emptyset\}$
\EndProcedure
\end{algorithmic}
\caption{Find indices of the right context to create a branch for. Further explanation in Figure \ref{fig:lex-right-quotient}.}
\label{algorithm:calc_boundaries_of_right_context}
\end{figure}

\begin{figure}
    \centering
    \includegraphics[width=\linewidth,page=2]{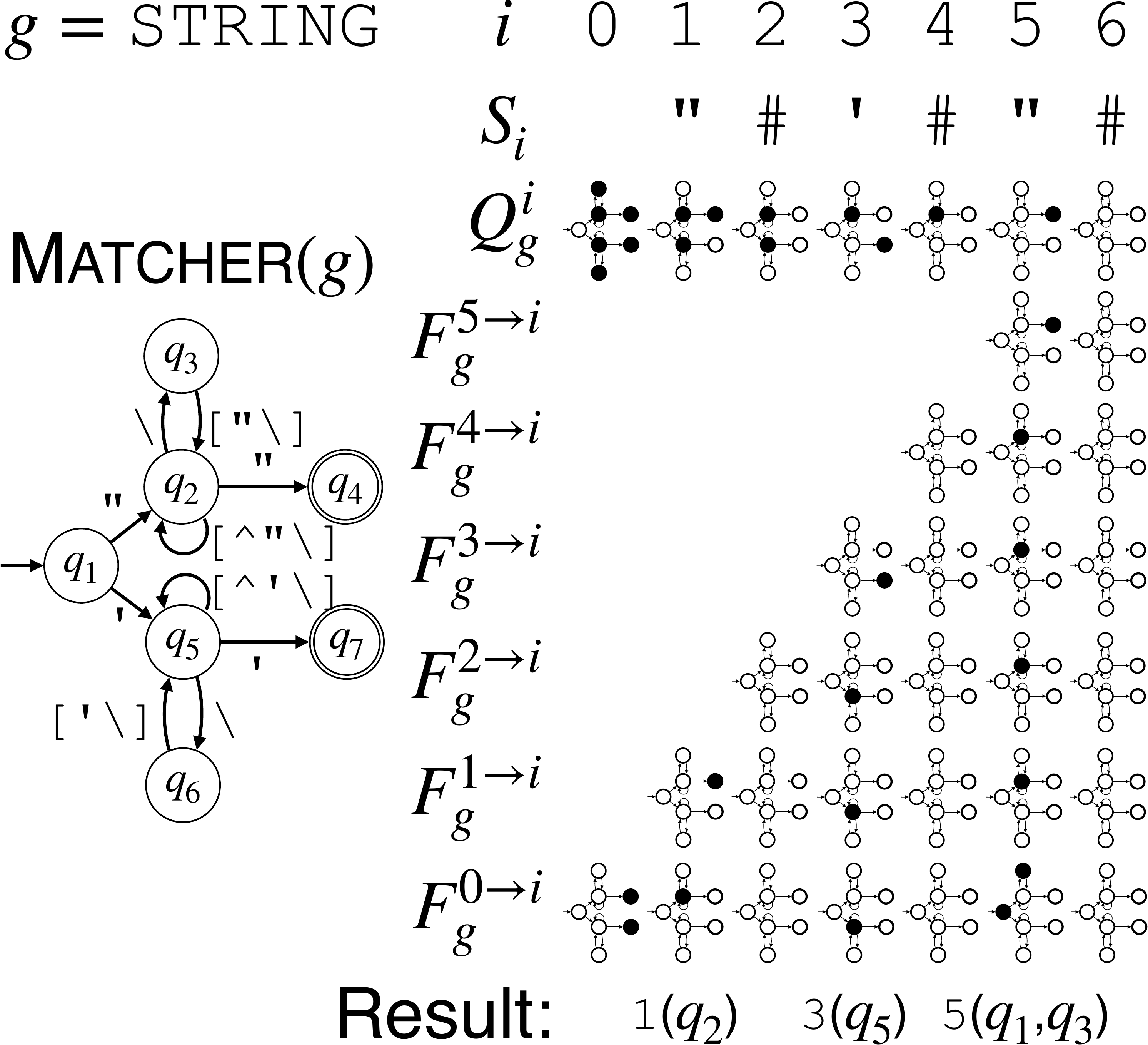}
    \caption{Illustration of algorithm from Figure \ref{algorithm:calc_boundaries_of_right_context} for the \texttt{STRING} lexeme, showing that it is valid to begin lexing after 1, 3, and 5 characters if the quotient text includes an unclosed string.
    Additionally, the algorithm provides the NFA states that should be active at the end of the quotient text, to connect to the right context.
    For each $i$, $Q^i_g$ represents the union, over all possible nonempty quotient texts $\alpha$, of the NFA states active after reading $\alpha \circ (S_1, \ldots, S_i)$.
    $F_g^{x \rightarrow i}$ represents, for each $x$, the possible NFA states that may be active at index $x$, that would cause a final state to be active at index $i$.
    Note that because we are only interested in longest matches, we remove any states from $F_g^{x \rightarrow i}$ that exist in $F_g^{y \rightarrow i}$ for $y > x$ (line \ref{line:subsumed_check}); this means that the size of a sparse representation of $F_g$ never exceeds $|\nfaStateSpace|$ at each step, so the runtime is $O(n)$.}
    \label{fig:lex-right-quotient}
\end{figure}

\section{Whitespace-Sensitive Languages}
\label{subsec:whitespace-sensitive}
Several programming languages, including Python, Haskell, and F\#, are sensitive to whitespace.
In Python, for example, indentation is used to delineate blocks of code.

The parser itself does not handle indentation; rather, this feature is implemented in the lexer, which emits special \texttt{INDENT} and \texttt{DEDENT} symbols according to a straightforward algorithm:
When the first non-whitespace character on a line is encountered, other than comment-only lines, the lexer counts the number of spaces---the \emph{indentation level} of the line \cite{PythonLexicalAnalysis}.

If the indentation level of the current line is greater than that of the previous line, the previous indentation level is added to a stack, and the lexer emits \texttt{INDENT} after the newline.
Alternatively, if the current line's indentation level is less than that of the previous line, the lexer pops indentation levels off of the stack until an indentation level matches; inserting \texttt{DEDENT} for each pop.
If no indentation level from the stack is matched, the lexer emits a syntax error.
Additionally, indentation is ignored after a line continuation character, or within parentheses.

This indentation rule does not cause too many interesting issues for an incremental parser; because the indentation level is not calculated for whitespace-only lines, there is always some valid completion---additional whitespace if the indentation level is too low, or a newline if too high. 
The lexer tracks the parentheses nesting depth in a straightforward manner.

\subsection{Right-quotienting with whitespace sensitivity}

\begin{figure}
 \centering
    \begin{subfigure}[t]{\linewidth}
    \begin{minted}[escapeinside=||]{python}
if foo:      |\highlightCodeBlue{IF NAME COLON}|
    if bar:  |\highlightCodeBlue{NL INDENT IF NAME COLON}|
        pass |\highlightCodeBlue{NL INDENT PASS}|
    pass     |\highlightCodeBlue{NL \highlightCode{DEDENT} PASS}|
             |\highlightCodeBlue{NL DEDENT}|
    \end{minted}
    \end{subfigure}
    \begin{subfigure}[t]{\linewidth}
    \begin{minted}[escapeinside=||]{python}
if foo:      |\highlightCodeBlue{IF NAME COLON}|
  if bar:    |\highlightCodeBlue{NL INDENT IF NAME COLON}|
    pass     |\highlightCodeBlue{NL \highlightCodeRed{INDENT} PASS}|
             |\highlightCodeBlue{NL DEDENT \highlightCodeRed{DEDENT}}|
    \end{minted}
    \end{subfigure}
    \caption{Despite having identical text on the last two lines, the lexer's output is significantly different.}
    \label{fig:ambiguous-dedents}
\end{figure}

Whitespace sensitivity significantly complicates the right quotienting operation.
Figure \ref{fig:ambiguous-dedents} shows how the right context, ``\verb|\n    pass\n|'', may correspond to multiple streams of symbols, depending on an unknown prefix.
Even this simple example may lead to many different parses---the second-to-last line may have an arbitrary number of \verb!DEDENT! symbols, if the preceding text is highly indented.
We resolve this by describing the set of possible lexer outputs as a regular language: \verb!NL (INDENT | DEDENT*)! \verb! PASS NL DEDENT{1,4}!.
Our method constructs this language, and uses the algorithm from Section \ref{section:earley_parsing} to compute a quotient grammar that will accept \emph{any} string which matches the right context's indentation.

There are several modifications to the indentation algorithm.
When encountering the first newline of the right context, the previous indentation level is completely unknown; so the lexer emits \texttt{(INDENT | DEDENT*)}.
Second, when the lowest indentation level seen so far in the right context is $n$, and the next indentation level is $m < n$, the lexer emits \verb!DEDENT{1,n-m}!, as there could be an indentation level at every step. 
The lexer keeps track of all ``lowest indentation level'' values of $m$; in the quotient language, $m$ must be on the indentation stack at the end when lexing a given string.

\subsection{A Completeness-Soundness-Complexity Tradeoff}

\label{subsec:indentation-flaw}

This procedure is \textit{complete}---it accepts all valid programs. However, it is not \textit{sound}, as some invalid programs are accepted. 
Figure \ref{fig:mismatched-indentation} provides two examples in which this occurs.

This issue stems from the fact that indentation lengths are variable.
It is possible to trade completeness for soundness; for example, by assuming all indentation is a fixed length---four spaces.
However, this assumption is violated in a significant portion of existing code.

It is also possible to trade both completeness and soundness for complexity.
For example, when decreasing the indentation level by four, instead of generating a \verb!DEDENT{1,4}!, the lexer can emit four separate branches, utilizing the branching infrastructure from Section \ref{section:lcfls}.
However, this approach is computationally intractable.

In practice, an engineer will need to decide on a suitable balance.
We present our solution as a reasonable default.

First, our program examines the given context; if the indentation always increases by a fixed amount, then our lexer operates in a strict mode; except for the initial newline, the number of \verb|DEDENT| symbols is always known exactly.
Otherwise, the lexer continues to issue variable \verb|DEDENT|s.

Regardless of this strict mode, the lexer does not ``know'' if the first newline of the right context represents an \verb|INDENT|, or any number of \verb|DEDENT|s.
We accept a small increase in complexity by creating two branches; one with \verb|INDENT| after the first newline, and the other with \verb|DEDENT*|. 
In the former branch, a string is only accepted as a member of the quotient language if its final indentation level is less than the indentation level after the first newline of the right context; the opposite is true for the latter branch.
Of course, it is possible to further increase the complexity by introducing additional branches---for example, by breaking out the \verb!DEDENT*! case---but we find that this approach is a useful point in the tradeoff.

\begin{figure}
 \centering
    \begin{subfigure}[t]{\linewidth}
    \begin{minted}[escapeinside=||]{python}
|\highlightCode{if foo:     }|  |\highlightCodeBlue{IF NAME COLON}|
|\highlightCode{     if foo:}|  |\highlightCodeBlue{NL INDENT IF NAME COLON}|
    pass      |\highlightCodeBlue{NL (\highlightCodeRed{INDENT}\pipe{}DEDENT*) PASS}|
              |\highlightCodeBlue{NL DEDENT\{1,4\}}|
    \end{minted}
    \end{subfigure}
    \begin{subfigure}[t]{\linewidth}
    \begin{minted}[escapeinside=||]{python}
|\highlightCode{if foo:      }| |\highlightCodeBlue{IF NAME COLON}|
|\highlightCode{  try bar:   }| |\highlightCodeBlue{NL INDENT TRY NAME COLON}|
|\highlightCode{    if baz:}|   |\highlightCodeBlue{NL INDENT IF NAME COLON}|
      pass    |\highlightCodeBlue{NL (INDENT\pipe{}DEDENT*) PASS}|
    except e: |\highlightCodeBlue{NL DEDENT\{1,\highlightCodeRed{2}\} EXCEPT...}|
      pass    |\highlightCodeBlue{NL INDENT PASS}|
              |\highlightCodeBlue{NL DEDENT\{\highlightCodeRed{2},5\}}|
    \end{minted}
    \end{subfigure}
    \caption{Two cases where a given quotient text incorrectly matches the quotient language, without strict lexing. (Top) On line 3, the \texttt{INDENT} is part of the right context's language, even though no indent is actually taken. (Bottom) The \texttt{DEDENT\{1,2\}} on line 5 allows for the case where both indentation levels 4 and 5 exist in the quotient language, and the \texttt{DEDENT\{2,5\}} on line 7 allows for the case where there are no indentation levels between 0 and 4. Because the indentation lengths are variable, this combination allows code where the \texttt{if} statement lines up with an \texttt{except} clause.}
    \label{fig:mismatched-indentation}
\end{figure}

\subsection{Parentheses in the Right Context}

As previously mentioned, parentheses affect the indentation of a Python program.
One seemingly-simple approach to handling parentheses is to count the number of closing parentheses in the right context, and subtract the number of opening parentheses, thus obtaining the parentheses-nesting level at the insertion point.
However, the implementation can be tricky: because parentheses in comments and strings are excluded, this count requires an additional full lexer pass.

We instead rely on a now-familiar pattern: the lexer ``guesses'' the parentheses level by creating two branches: one where the initial parentheses level is zero, and one where it is nonzero.
In the first branch, newlines cause indentation to be processed; in the second branch, it is not.
Upon encountering an extra close-parentheses, the first branch is removed, and the second branch splits further into two branches.
Finally, when done parsing the right context, any branch where the current nesting level is not zero is removed.
The lexer maintains a count of how much the nesting level has changed since the beginning of the right context; when checking for quotient language membership, the current nesting level must match.

\section{Experiments}

\label{section:experiments}

Using the methods described so far in this paper, we implemented\footnote{Link omitted for review; see supplemental material.} a proof-of-concept incremental right-quotienting parser for Python 3.
We evaluate the effectiveness of constrained generation on a LLM using our parser, through a simulated code completion task.

\subsection{Datasets}

The Stack \cite{kocetkovStackTBPermissively2022} is a large dataset of publicly-available code sourced from GitHub.
For our evaluations, we use a subset of this dataset (``\verb|the-stack-smol-xl|''), which contains 10,000 randomly selected Python files.
We exclude 459 files for which \verb|ast.parse| in Python 3.9 returns an error (mostly Python 2 files), and 2 files which use features we did not implement;\footnote{One file mixes tabs and spaces in an unusual way; the other exhibits a rare edge case of tuple unpacking in certain contexts that was not accounted for in the CFG included in the Lark package \cite{shinanerezLarkParsingToolkit}, which our grammar is based on. Note that Python is specified as a PEG, so exact conversion to a CFG would be nontrivial or impossible, and as such was not a goal.} this leaves 9539 files.

From this base dataset, we construct two synthetic datasets for the code completion task.
The first dataset, \textsc{Stack-Boundary}, is constructed by running the Python tokenizer on the original file, and selecting a pair of symbols at the same indentation level.
The text between these two symbols is removed, along with a random amount of the first symbol, to simulate the common task of complete-as-you-type within an IDE.
This is done for 10 random symbol pairs for each file, obtaining 95390 experiments.
We include the code and random seeds necessary to exactly reproduce both datasets in our supplemental material.

The second dataset, \textsc{Stack-RandSpan}, uses random span deletions, similar to the random span infilling benchmark presented in \cite{bavarianEfficientTrainingLanguage2022}, allowing us to validate the performance of our method in more technically challenging mid-symbol scenarios.
For each file, we choose a random point up to 90\% of the way through the file, and delete up to 100 characters, 20\% of the file, or until the end of the file, whichever comes first.
This is also done at 10 random insertion points for each file.

\subsection{Experimental Setup}

For each experiment, we use SantaCoder \cite{allalSantaCoderDonReach2023} with a FIM task to generate code.

We first obtain the quotient language of Python with respect to the left and right contexts.
We then construct the input to the LLM, as described in Section \ref{section:fim-special-tokens}.
Note that the input to the LLM may not be identical to the context used to create the quotient language.
For example, if the context is larger than the LLM's maximum input size, the beginning of the left context and end of the right context may be truncated when constructing the input.
Conversely, more sophisticated LLM systems may receive an input that spans several files, but the quotient language relies only on the context of one file.

We utilize greedy sampling to choose tokens; if the end of sequence token has not been generated after 500 tokens, we stop generation.
We use the incremental parsability check on the quotient language for the highest-probability token; if it fails, the program tests the next one, up to the top 50 token candidates.
If the highest-priority candidate is \textsc{EOS}, we instead check that the text generated so far is a member of the quotient language.
If all 50 candidates fail, we stop generation, as if the token limit were reached.
Occasionally, the LLM generates a sequence that contains valid stopping points, but never selects \textsc{eos} prior to the token limit.
In this case, we select the token index with the highest \textsc{eos} probability (weighted by generation length) for which the language membership test succeeds.
If none exist, our method reports failure.

As discussed in Section \ref{subsection:checked-unconstrained-why-not}, it is possible to simply run unconstrained generation, and check each prefix for parsability in context using \texttt{ast.parse}.
This method, checked unconstrained generation, will succeed on a superset of cases compared to regular unconstrained generation.

Finally, for all of these methods, we consider a generation to be successful if \texttt{ast.parse} succeeds on the result.

\subsection{Results \& Discussion}

\begin{table}
  \begin{tabular}{c|cc|c}
    \toprule
    &Constrained \cmark& Constrained  \xmark\ & Total \\
    \midrule
    Unconstrained \cmark &  65342 & 11 & 65353\\
    Unconstrained \xmark & 24833 & 94 & 24927 \\
    Checked Unconstrained \xmark & 490 & 4620 & 5110\\
    \midrule
    Total & 90665 & 4725 & 95390 \\
  \bottomrule
\end{tabular}
  \caption{Results for \textsc{Stack-Boundary}}
    \label{tab:results-boundary}
\end{table}

\quad 

\begin{table}
  \begin{tabular}{c|cc|c}
    \toprule
    &Constrained  \cmark & Constrained  \xmark & Total \\
    \midrule
    Unconstrained \cmark & 68570 & 20 & 68590\\
    Unconstrained \xmark & 22952 & 112 & 23064\\
    Checked Unconstrained \xmark & 563 & 3173 & 3736\\
    \midrule
    Total & 92085 & 3305 & 95390 \\
  \bottomrule
\end{tabular}
  \caption{Results for \textsc{Stack-RandSpan}}
  \label{tab:results-randspan}
\end{table}

We include the results of our experiments in Tables \ref{tab:results-boundary} and \ref{tab:results-randspan}. 
With both datasets, our method performs significantly better than unconstrained generation.
Checked unconstrained generation fixes cases where the unconstrained model never generates \textsc{eos}, but it sometimes generates code for which there can never be a valid completion, so constrained generation still performs slightly better.
Surprisingly, all methods perform better on \textsc{Stack-RandSpan} than on \textsc{Stack-Boundary}. 
We hypothesize that this is caused by \textsc{Stack-RandSpan} having less text removed, on average, despite containing more technically challenging right contexts; in an early version of our experiments where the \textsc{Stack-RandSpan} removals were not capped at 100 characters, all methods performed worse.

\subsubsection{Failure Analysis}
Of the 4725 unsuccessful generation examples of \textsc{Stack-Boundary}, we distinguish between two cases.
In 29 cases, the incremental parser identified a left context as being in the quotient language, but the Python parser rejected the generated text.
In only one of these cases does unconstrained generation succeed; checked unconstrained generation succeeds on a further 22 instances.
These cases could be mitigated through further refinement of our incremental parser; though we note that doing so may require a complexity tradeoff as described in Section \ref{subsec:indentation-flaw}---a significant portion of these cases are caused by such indentation tradeoffs, or are due to certain exotic types of strings, for which we used a simplified implementation.

In the remaining 4696 cases, the code generation model doesn't succesfully ``connect'' to the right context before generation is cut off; for example, by continuously producing import statements.
Unconstrained generation succeeds in 10 of these cases, and checked unconstrained generation in a further 72 cases.
These could potentially be mitigated through a method of ``steering'' generation towards finite branches in the grammar; this is an area for future work.
This number also includes cases where the model is forced into a low-probability branch, resulting in the only syntactically-valid completions being outside of the top 50 tokens. 
Such cases can be mitigated by layering our method with the technique presented in \cite{park2024grammaraligned}.

\subsection{Runtime Performance}

\begin{figure}
    \centering
    \includegraphics[width=0.4\textwidth]{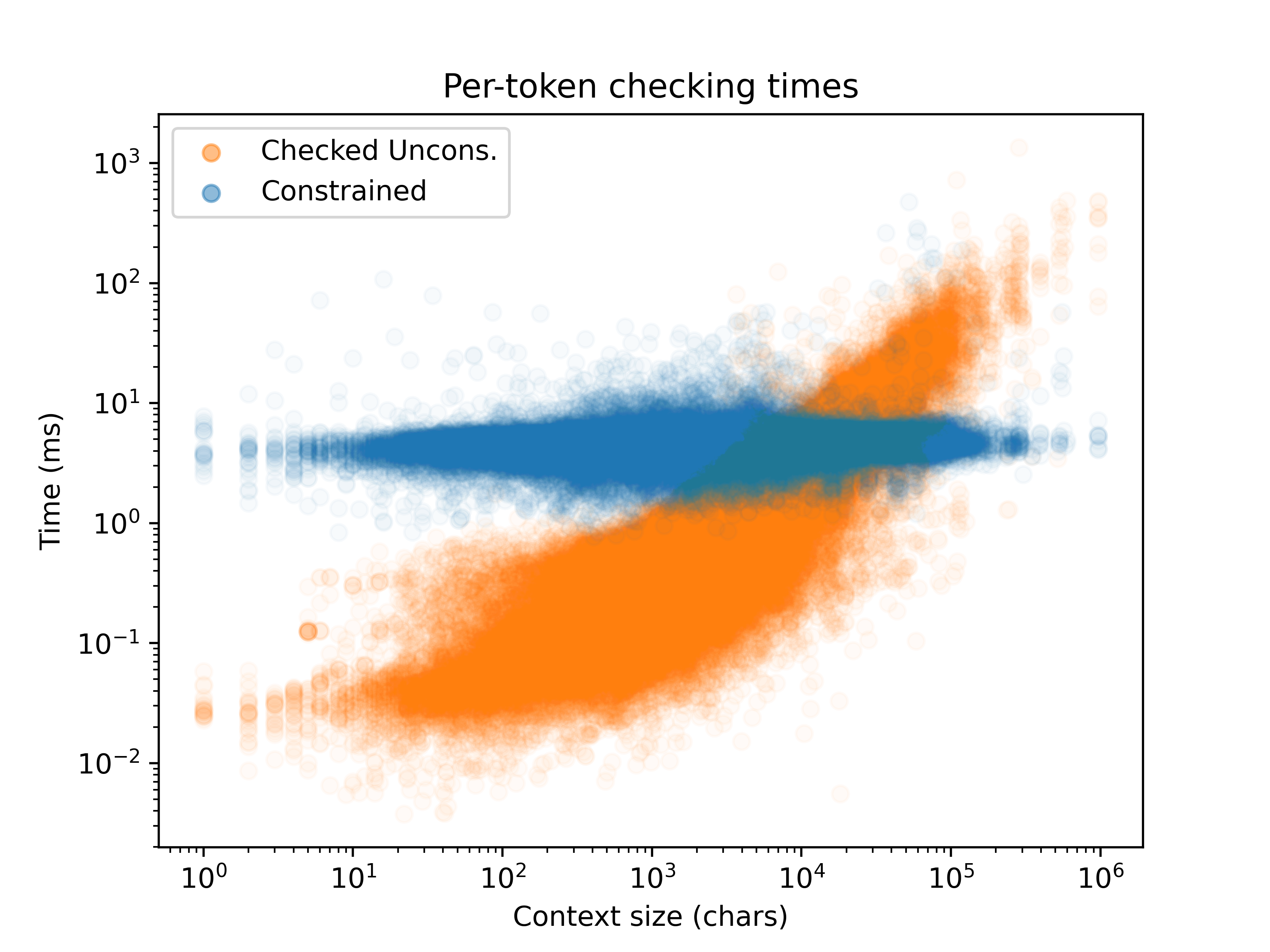}

    \includegraphics[width=0.4\textwidth]{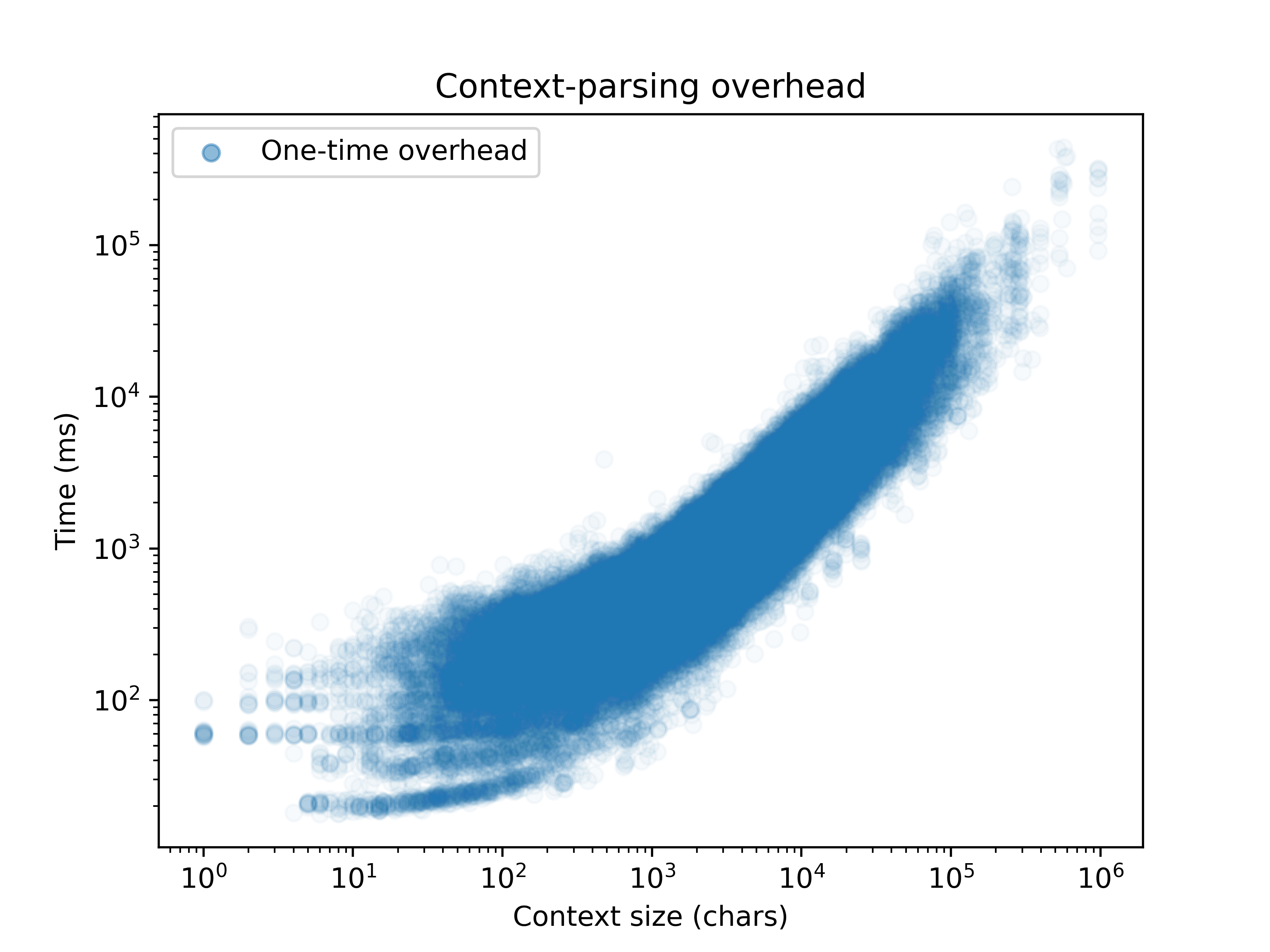}
    
    \caption{Timing graphs of our prototype. Note that line of best fit is calculated but not displayed due to log-log axes. \\(Top) Mean overhead per generated token. \\Checked uncons.: $y = (3.7\times 10^{-4}) x - 0.177$ ($R^2 = 0.544$). Constrained: $y=(9.54\times 10^{-6}) x + 4.53$ ($R^2 = 4.22\times 10^{-3}$). \\(Bottom) One-time overhead for constrained generation. \\$y = 0.283x + 270$ ($R^2 = 0.729$).}
    \label{fig:per-token-times}
\end{figure}

It is important to note that our implementation is a research prototype, written largely in Python, with selected subroutines written in Rust.
Therefore, the performance results we present will differ considerably compared to a production-quality implementation, likely by several orders of magnitude.

As mentioned in Section \ref{subsection:checked-unconstrained-why-not}, checked unconstrained generation requires parsing the entire file for every new token, leading to undesirable asymptotic complexity.
As Figure \ref{fig:per-token-times} (top) shows, our method is able to parse each new token more quickly at longer context lengths, as per-token generation time is almost completely uncorrelated with context size.
The main overhead, shown in Figure \ref{fig:per-token-times} (bottom), only occurs once per generation.
Our proof-of-concept prototype is already competitive on performance compared to \texttt{ast.parse} for long or repeated generations at large context sizes (context size ${\sim}10^4$ for per-token generation times to be faster with constrained generation, with ${\sim} 10^3$ tokens generated to overcome the one-time context parsing overhead); a tuned and production ready implementation of this method would likely be orders of magnitude faster than the prototype.
A production system may also intelligently select a method depending on context size, generation lengths and repetitions, and tolerance for errors.

\section{Future Work}
\label{section:discussion}

As discussed previously, one area of future work is in ``steering'' code generation to complete a grammar, rather than constraining generation to be a valid prefix.
This task is similar to the problem presented in \cite{huImprovedLexicallyConstrained2019}, where specific tokens must be included in a LLM's output.
We present two additional areas for future work:

\subsection{Preventing Context Escape}

While the constrained generation system effectively generates valid code, it still allows code which may not be desirable in an interactive completion setting.
For example, if a user places their cursor within the parentheses of a function call, it can usually be inferred that the intention is to add or edit a parameter, without exiting the parameter list.
However, it is still syntactically valid for the LLM to generate a close parentheses, construct a newline, output the name of another function, and finally generate an open parentheses.
We refer to this phenomenon as \emph{Context Escape}.

While more complete code generation systems, and evaluations for systems that include metrics of context escape, are out of scope for this paper, we believe that the methods we have presented could be useful to create systems that handle context escape.
In particular, the explicit CFG produced by the quotienting operation provides a promising structure on which to apply language-specific heuristics.

\balance
\subsection{Static Analyses Beyond Well-Formedness}

\label{subsec:other_static_analyses}

This paper presents lexing and parsing infrastructure for partial programs that contain a single insertion point.
However, well-formedness is just one static property of a program.

A natural question is whether we can use left and right quotienting operations to statically analyze other properties of a partial program: type and memory safety, exclusion of hallucinated or memorized PII and credentials, proper sanitization of inputs, avoidance of biased behavior based on a person's protected class, and so on.
We leave these type of static analyses, and their incorporation into a larger code generation system, as future work.

\section{Conclusion}
\label{section:conclusion}

In this work, we have demonstrated that a modification to the Earley parsing algorithm can be used to incrementally parse a quotient language.
We have introduced extensions to allow for context-sensitive features present in many programming languages, such as leftmost-longest lexing and whitespace sensitivity.
Finally, in our experiments, we integrate this algorithm into a Python 3 constrained generation system, enabling a LLM to produce syntactically correct code at a higher rate than with unconstrained generation on real-world codebases.

\bibliography{IEEEabrv,bibliography}

\end{document}